\overfullrule=0pt
\input harvmac

\def\p{\partial}
\def\pb{\overline\partial}
\def\O{\Omega}
\def\a{\alpha}
\def\b{\beta}
\def\g{\gamma}
\def\d{\delta}
\def\s{\sigma}
\def\r{\rho}
\def\l{\lambda}
\def\o{\omega}
\def\lb{\overline\l}
\def\ob{\overline\o}
\def\th{\theta}
\def\L{\Lambda}
\def\Lb{\overline\L}
\def\Lt{\widetilde\L}

\Title{\vbox{}} {\vbox{ \centerline{\bf The b Ghost of the Pure Spinor Formalism is Nilpotent}}}
\bigskip\centerline{Osvaldo Chandia\foot{e-mail: osvaldo.chandia@uai.cl}}
\bigskip
\centerline{\it Departamento de Ciencias, Facultad de Artes Liberales}
\centerline{\it Facultad de Ingenieria y Ciencias}
\centerline{\it Universidad Adolfo Iba\~nez, Santiago, Chile}

\vskip .3in

The ghost for world-sheet reparametrization invariance is not a fundamental field in the pure spinor formalism. It is written as a combination of pure spinor variables which have conformal dimension two and such that it commutes with the BRST operator to give the world-sheet stress tensor. We show that the ghost variable defined in this way is nilpotent since the OPE of $b$ with itself does not have singularities.

\vskip .3in

\Date{September 2010}

\lref\BerkovitsFE{
  N.~Berkovits,
  ``Super-Poincare Covariant Quantization of the Superstring,''
  JHEP 0004 (2000) 018
  [arXiv:hep-th/0001035].}

\lref\BerkovitsNN{
  N.~Berkovits,
  ``Cohomology in the Pure Spinor Formalism for the Superstring,''
  JHEP 0009 (2000) 046
  [arXiv:hep-th/0006003].}

\lref\BerkovitsQX{
  N.~Berkovits and O.~Chandia,
  ``Massive Superstring Vertex Operator in D = 10 Superspace,''
  JHEP 0208 (2002) 040
  [arXiv:hep-th/0204121].}

\lref\BerkovitsPH{
  N.~Berkovits and B.~C.~Vallilo,
  ``Consistency of Super-Poincare Covariant Superstring Tree Amplitudes,''
  JHEP 0007 (2000) 015
  [arXiv:hep-th/0004171].}

\lref\BerkovitsPX{
  N.~Berkovits,
  ``Multiloop Amplitudes and Vanishing Theorems Using the Pure Spinor
  Formalism for the Superstring,''
  JHEP 0409 (2004) 047
  [arXiv:hep-th/0406055].}

\lref\BerkovitsBT{
  N.~Berkovits,
  ``Pure spinor formalism as an N = 2 topological string,''
  JHEP 0510 (2005) 089
  [arXiv:hep-th/0509120].}

\lref\OdaAK{
  I.~Oda and M.~Tonin,
  ``Y-formalism and $b$ Ghost in the Non-minimal Pure Spinor Formalism of Superstrings,''
  Nucl.\ Phys.\  B {\bf 779}, 63 (2007)
  [arXiv:0704.1219 [hep-th]].}

\lref\BerkovitsUS{
  N.~Berkovits,
  ``Relating the RNS and Pure Spinor Formalisms for the Superstring,''
  JHEP 0108 (2001) 026
  [arXiv:hep-th/0104247].}

\lref\SiegelXJ{
  W.~Siegel,
  ``Classical Superstring Mechanics,''
  Nucl.\ Phys.\  B {\bf 263}, 93 (1986).}

\newsec{Introduction}

The pure spinor formalism was invented ten years ago by Berkovits to provide a covariant quantization of the superstring \BerkovitsFE. The idea is to construct an action for the superstring which preserves spacetime supersymmetry and describes the superstring spectrum. In a flat background, the action depends on the ten-dimensional superspace variables and a bosonic pure spinor variable necessary for conformal invariance. Besides, Berkovits introduced a BRST charge which allows quantization. The spectrum coincides with the Green-Schwarz result in the light-cone gauge \BerkovitsNN. The spectrum can also be described in a ten-dimensional covariant manner \BerkovitsQX. Tree-level scattering amplitudes were defined in \BerkovitsPH\ where it was shown that this prescription reproduces certain RNS results.

A prescription to compute loops in scattering amplitudes was defined in \BerkovitsPX. It allows to show some vanishing theorems of supergravity effective actions. However complications coming from picture-changing operators can be avoided by adding new variables to the formalism without changing the physical content of the theory. This was done in \BerkovitsBT\ where the new variables also satisfy pure spinor conditions. This leads to the non-minimal pure spinor formalism that will be reviewed below.

The organization of this paper is as follows. In Section 2 we review the pure spinor formalism in the minimal and non-minimal versions. In section 3 we discuss general properties of the $b~b$ OPE and determine its vanishing by using cohomology arguments.

\newsec{Reviewing The Pure Spinor Formalism of the Superstring}

Consider a string in a flat ten-dimensional spacetime. The world-sheet action is given by
\eqn\action{S=\int d^2z ~ \ha~\p X^m \pb X_m + p_\a \pb\th^\a + \o_\a\pb\l^\a ,}
where $(X^m, \th^\a)$ are the coordinates of the $N=1$ superspace in ten dimensions\foot{That is, $m=0, \dots, 9$ and $\a=1, \dots, 16$.}, $p_\a$ is the canonical conjugate of $\th^\a$, the pure spinor $\l^\a$ is constrained to satisfy
\eqn\ps{(\l\g^m\l)=0,}
where $\gamma^m$ are the symmetric $16\times 16$ Pauli matrices in ten dimensions. The canonical conjugate $\o_\a$ of the pure spinor is defined up to
\eqn\ogauge{\d\o_\a=(\l\g_m)_\a\L^m .}
Note that \ps\ and \ogauge\ imply that $\l$ and $\o$ have eleven independent components respectively. This assures the conformal invariance of the action \action\ since the total central charge vanishes.

The quantization of this system is done after defining $Q=\oint\l^\a d_\a$ as BRST of the theory. Here, $d_\a$ is a supersymmetric combination of $X, p,\th$ given by
\eqn\dd{d_\a=p_\a-\ha (\g_m\th)_\a\p X^m-{1\over 8}(\g_m\th)_\a(\th\g^m\p\th) .}
Note that $Q^2=0$ is consequence of the OPE
\eqn\dd{d_\a(y) d_\b(z) \to -{{\g^m_{\a\b} \Pi_m}\over(y-z)} ,}
and the pure spinor condition \ps. Here $\Pi^m=\p X^m +\ha (\th\g^m\p\th)$ is the supersymmetric momentum. The OPE algebra among $(d_\a, \Pi^m, \p\th^\a)$ will be necessary below. To complete it, we need \SiegelXJ.
\eqn\opedpt{d_\a(y)\Pi^m(z)\to {{(\g^m\p\th)_\a}\over(y-z)}, \quad d_\a(y)\p\th^\b(z)\to {{\d_\a^\b}\over(y-z)^2}, \quad \Pi^m(y)\Pi^n(z)\to -{{\eta^{mn}}\over(y-z)^2} }

Although it is not known if $Q$ is a genuine BRST charge, in the sense that the gauge symmetry which is being fixed is unknown, we can study the cohomology of $Q$ and see if it has some physical meaning. It turns out that the cohomology of $Q$ is equivalent to the spectrum of the Green-Schwarz superstring as it was demonstrated in \BerkovitsNN\ using the light-cone gauge. In a covariant language, the physical states are described by ghost number one vertex operators. For massless states, the vertex operator $U=\l^\a A_\a(X,\th)$ satisfies the equations of motion and the gauge invariance of super-Maxwell system in ten dimensions after declaring that $U$ is in the cohomology of $Q$, that is $QU=0$ and $U\sim U+Q\O$. Similarly, the first massive state was discussed in \BerkovitsQX.

The pure spinor variables enter in invariant combinations under \ogauge. If we consider holomorphic combinations only, we have the possibilities $T_{\rm{pure}}=-\o_\a\p\l^\a$, which contributes to the stress tensor, $J=-\o_\a\l^\a$, which can be defined as the ghost number current, and $N^{mn}=\ha(\l\g^{mn}\o)$, which can be defined as the Lorentz current for the pure spinor variables. Of course we can define others combinations that are invariant under \ogauge, but they can be written in terms of $(T_{\rm{pure}}, J, N^{mn})$. We will need the OPE algebra between $J$ and $N^{mn}$, it is \BerkovitsUS\
\eqn\openj{J(y)J(z)\to -{4\over(y-z)^2}, \quad J(y) N^{mn}(z)\to 0 ,}
$$
N^{mn}(y)N^{pq}(z)\to {3\over(y-z)^2}\eta^{m[p}\eta^{q]n} - {1\over(y-z)} ( \eta^{m[p}N^{q]n}-\eta^{n[p}N^{q]m} ) .$$

To compute scattering amplitudes, the formalism needs to break manifest Lorentz invariance \BerkovitsPX. In order to preserve Lorentz covariance, it is necessary to modify the formalism without spoiling its properties. This is the non-minimal pure spinor formalism \BerkovitsBT. We add new variables. They were called non-minimal variables in \BerkovitsBT\ and are a bosonic spinor $\lb_\a$ and a fermionic spinor $r_\a$ constrained to satisfy
\eqn\nonmps{(\lb\g^m\lb)=0, \quad (\lb\g^m r)=0 .}
Note that these constraints imply that $\lb_\a$ and $r_\a$ have eleven independent components respectively. The action for them is of the form
\eqn\actionnonm{S_{nm}=\int d^2z ~ \ob^\a\pb\lb_\a + s^\a\pb r_\a ,}
where $\ob^\a$ is the canonical conjugate of $\lb_\a$ and $s^\a$ is the canonical conjugate of $r_\a$. Each conjugate variable has eleven independent components because the gauge invariance
\eqn\gaugenonmin{\d\ob^\a=(\g^m\lb)_\a \Lb_m-(\g^m r)^\a \Lt_m ,}
$$
\d s^\a=(\g^m\lb)^\a \Lt_m .$$

At this point we could ask if it is possible to construct a $b$ ghost. In the minimal version it is possible to write \BerkovitsUS\
\eqn\bnm{b={{C_\a G^\a}\over{C_\a\l^\a}} ,}
where $G^\a$ is
\eqn\gg{G^\a=\ha \Pi^m(\g_m d)^\a - {1\over 4} N^{mn}(\g_{mn}\p\th)^\a - {1\over 4} J \p\th^\a-  {1\over 4} \p^2\th^\a ,}
and satisfies $QG^\a=\l^\a T$ with $T$ being the stress tensor \BerkovitsUS. Since $C_\a$ is a constant spinor, the $b$ ghost defined in this way is clearly non covariant. In the next section we will review how to define a covariant $b$ ghost using the non-minimal variables. Before, we will study some general properties that can be learnt for an arbitrary $b$ ghost.

\newsec{Nilpotency of the $b$ Ghost}

In this section we will argue that the $b$ ghost of the non-minimal formalism is nilpotent. We will perform a general analysis first, by constraining the form of the $b~b$ OPE dictated from conformal invariance and BRST invariance.

\subsec{Generic analysis}

The b ghost comes from gauge fixing the parametrization invariance of a string. In the conformal gauge, it is a world-sheet field of conformal dimension two which is nilpotent and satisfies

\eqn\bghost{Qb=T,}
where $Q$ is the BRST charge and $T$ is the stress tensor.  The generic OPE of $b$ with itself has the form

\eqn\bb{b(y)b(z)\to {O_0(z)\over(y-z)^4}+ {O_1(z)\over(y-z)^3}+ {O_2(z)\over(y-z)^2}+ {O_3(z)\over(y-z)}, }
where $O_n$ is a field of conformal dimension $n$ which are not necessarily primary fields.

Now we are going to impose some restrictions on these fields. The first ones come from the Grassmannian nature of $b$, from which we know that $b(z)b(y)=-b(y)b(z)$ is satisfied. Using this and the expression \bb, we obtain

\eqn\bodd{O_0=0,\quad O_2=\ha\p O_1 .}
Up to now we have,

\eqn\bbb{b(y)b(z)\to {O_1(z)\over(y-z)^3}+\ha {\p O_1(z)\over(y-z)^2}+ {O_3(z)\over(y-z)}.}
Now, we will apply the stress tensor $T(w)$ to this expression and we use the fact that the $b$ ghost is a primary field \OdaAK. On the l.h.s. we obtain

$$
T(w)b(y)b(z)\to({2b(y)\over(w-y)^2}+{\p b(y)\over(w-y)})b(z)+b(y)({2b(z)\over(w-z)^2}+{\p b(z)\over(w-z)}),$$
and now we use \bbb\ to get

$$
{2\over(w-y)^2}({O_1(z)\over(y-z)^3}+\ha{\p O_1(z)\over(y-z)^2}+{O_3(z)\over(y-z)})$$
$$ + {1\over(w-y)}({-3O_1(z)\over(y-z)^4}-{\p O_1(z)\over(y-z)^3}-{O_3(z)\over(y-z)^2})
+{2\over(w-z)^2}({O_1(z)\over(y-z)^3}+\ha{\p O_1(z)\over(y-z)^2}+{O_3(z)\over(y-z)})$$
$$
+{1\over(w-z)}({3O_1(z)\over(y-z)^4}+{2\p O_1(z)\over(y-z)^3}+\ha{\p^2O_1(z)\over(y-z)^2}+{O_3(z)\over(y-z)^2}+{\p O_3(z)\over(y-z)}) .$$
We organize this in terms of expansion in inverse powers of $(y-z)$ and inverse powers of $(w-z)$. To do this we need to use

$$
{1\over(w-y)}={1\over(w-z)}+{1\over(w-z)^2}(y-z)+{1\over(w-z)^3}(y-z)^2+{1\over(w-z)^4}(y-z)^3 + {\cal O}((y-z)^4),$$
$$
{1\over(w-y)^2}={1\over(w-z)^2}+{2\over(w-z)^3}(y-z)+{3\over(w-z)^4}(y-z)^2+ {\cal O}((y-z)^3),$$
to obtain

$$
{1\over(y-z)^3}({O_1(z)\over(w-z)^2}+{\p O_1(z)\over(w-z)})+
{1\over(y-z)^2}({O_1(z)\over(w-z)^3}+{\p O_1(z)\over(w-z)^2}+\ha{\p^2 O_1(z)\over(w-z)})$$
$$
+{1\over(y-z)}({3O_1(z)\over(w-z)^4}+{\p O_1(z)\over(w-z)^3}+{3O_3(z)\over(w-z)^2}+{\p O_3(z)\over(w-z)}) .$$
Applying $T(w)$ on the r.h.s. of \bbb, and comparing with the above expression we determine the OPE's between the stress tensor and the $O$ operators. We obtain

\eqn\toum{
T(w) O_1(z) \to {O_1(z)\over(w-z)^2}+{\p O_1(z)\over(w-z)},}
which implies that $O_1$ is a primary field,

$$
\ha T(w) \p O_1(z) \to {O_1(z)\over(w-z)^3}+{\p O_1(z)\over(w-z)^2}+\ha{\p^2 O_1(z)\over(w-z)},$$
which is consistent with the $T O_1$ OPE, and

\eqn\totres{
T(w) O_3(z) \to {3O_1(z)\over(w-z)^4}+{\p O_1(z)\over(w-z)^3}+{3O_3(z)\over(w-z)^2}+{\p O_3(z)\over(w-z)},}
which states that $O_3$ is not primary.

Consider again \bbb. Now we act this expression with $Q$ and use the fact that the $b$ ghost is primary. On the l.h.s we obtain

$$
Qb(y) b(z) - b(y) Qb(z) = T(y) b(z) - T(z) b(y) \to 0 .$$
On the r.h.s, this implies that $Q O_1=Q O_3 = 0$.

\subsec{The pure spinor case}

Now, we consider  the BRST charge and the $b$ ghost for the pure spinor in its non-minimal version. The BRST charge is given by
\eqn\Q{Q=\oint \l^\a d_\a + \ob^\a r_\a ,}
and the $b$ ghost is given by
\eqn\bnm{b=b_{-1}+b_0+b_1+b_2+b_3 }
$$
=-s^\a \p\lb_\a + {1\over(\l\lb)} \lb_\a G^\a - {1\over(\l\lb)^2} \lb_\a r_\b H^{\a\b} - {1\over(\l\lb)^3} \lb_\a r_\b r_\g K^{\a\b\g} + {1\over(\l\lb)^4} \lb_\a r_\b r_\g r_\d L^{\a\b\g\d} ,$$
where $(G, H, K, L)$ are conformal dimension $2$ fields which satisfies
\eqn\qghkl{QG^\a=\l^\a T,\quad QH^{\a\b}=\l^{[\a}G^{\b]},\quad QK^{\a\b\g}=\l^{[\a}H^{\b\g]},\quad QL^{\a\b\g\d}=\l^{[\a}K^{\b\g\b\d]},\quad \l^{[\a}L^{\b\g\d\r]}=0.}
These fields depend on the minimal fields in the form
\eqn\G{G^\a=\ha \Pi^m(\g_m d)^\a - {1\over 4} N^{mn}(\g_{mn}\p\th)^\a - {1\over 4} J \p\th^\a-  {1\over 4} \p^2\th^\a ,}
\eqn\H{H^{\a\b}={1\over 192} \g^{\a\b}_{mnp} ( (d\g^{mnp}d) + 24 N^{mn} \Pi^p ) ,}
\eqn\K{K^{\a\b\g}={1\over 16}\g_{mnp}^{[\a\b}(\g^md)^{\g]}N^{np} ,}
\eqn\L{L^{\a\b\g\d}={1\over 128}\g_{mnp}^{[\a\b} (\g^{pqr})^{\g\d]} N^{mn} N_{qr} .}

Note that when the triple pole in the $b(y)b(z)$ OPE vanishes, \bbb\ will be of the form
\eqn\BB{b(y)b(z)\to {O_3(z)\over(y-z)} ,}
where $O_3$ is a primary conformal dimension $3$ field which is annihilated by  $Q$. Cohomology arguments will imply that $O_3$ is BRST trivial. Moreover, we will show that $O_3$ vanishes if we demand supersymmetry covariance and use cohomological arguments due to the form of the BRST charge \Q.

The OPE $b(y)b(z)$ is an expansion in powers of $r_\a$. The term $r^n$ comes from contractions between $b_k$ with $b_{n-k}$ with $k=0,1,2,3$. Now we will show that the triple pole in this OPE vanishes order by order in powers of $r$ because the pure spinor conditions for the non-minimal variables \nonmps.

At order $r^0$ ,  we consider the OPE's  $b_{-1}(y) b_1(z) - (y \leftrightarrow z)$  and $b_0(y)b_0(z)$. Only the last one will produce a triple pole. In fact, it comes from the OPE between the term $\Pi^m(\g_m d)^\a$ in \G\ with itself. It is of the form
$$
\Pi^m(\g_m d)^\a(y)\Pi^n(\g_n d)^\b(z) \to {1\over(y-z)^3} \g_m^{\a\b} \Pi_m(z) + \cdots .$$
Therefore, it vanishes after hitting  with $\lb_\a\lb_\b/(\l\lb)^2$ because the pure spinor conditions \nonmps.


At order $r^1$ , we consider the OPE's  $b_{-1}(y)b_2(z) - (y \leftrightarrow z)$ and $b_0(y)b_1(z)- (y \leftrightarrow z)$. The last OPE produces a triple pole. It comes from contractions between first term in \G\ with the first term in \H. We obtain an OPE of the form
$$
G^\a(y) H^{\b\g}(z) \to {1\over(y-z)^3} \g_{mnp}^{\b\g} (\g^{mnp}d)^\a + \cdots .$$
After hitting with $\lb_\a\lb_\b /(\l\lb)^3$ we obtain that this singularity is proportional to
$$
(\lb\g^{mnp}d)(\lb\g_{mnp}r) ,$$
which vanishes after factoring out $\lb_\a\lb_\b$, symmetrizing in $(\a\b)$ and using the identity for the gamma matrices
\eqn\Fierz{
(\g^m)^{(\a\b}(\g_m)^{\g)\d}=0.}

At order $r^2$ , we consider the OPE's  $b_{-1}(y)b_3(z) - (y \leftrightarrow z)$, $b_0(y)b_2(z)- (y \leftrightarrow z)$ and $b_1(y) b_1(z)$. The last two possibilities determine triple pole singularities. Consider first $b_0(y)b_2(z)- (y \leftrightarrow z)$.  The triple pole comes from contractions between the second term in \G\ with \K. We also obtain a fourth pole. The result is
$$
G^\a(y) K^{\b\g\d}(z) \to {1\over(y-z)^4} (\g_{mn}\g_p)^{\a[\b} (\g^{mnp})^{\g\d]} + {1\over(y-z)^3} (\g_{mn}\g_p)^{\a[\b}(\g^{mp}{}_q)^{\g\d]} N^{nq}$$
$$
+ {1\over(y-z)^3} (\g^m)^{\a[\b} \g_{mnp}^{\g\d]} N^{np} + \cdots ,$$
which vanishes after hitting with $\lb_\a \lb_\b r_\g r_\d$ because of the pure spinor conditions. Note that the fourth pole vanishes, but it potentially implies a third pole singularity since we need to expand $\lb(y)$ around $z$. It also vanishes because $(\p\lb\g^m\lb)=0$. Consider now $b_1(y) b_1(z)$. The triple and higher poles come from contractions between the first term in \H\ with itself and between the second term in \H\ with itself. After hitting with $\lb_\a r_\b/(\l\lb)^2(y)\lb_\g r_\d/(\l\lb)^2$ we obtain
$$
{1\over(y-z)^4} (\lb\g_{mnp}r)(y) (\lb\g^{mnp}r)(z) + {1\over(y-z)^3} (\lb\g_{mnp}r)(y) (\lb\g^{mn}{}_q r) N^{pq}(z) + \cdots ,$$
which vanishes because of \Fierz.

At order $r^3$ , we consider the OPE's $b_0(y)b_3(z) - (y \leftrightarrow z)$ and $b_1(y)b_2(z)- (y \leftrightarrow z)$. Consider first $b_0(y)b_3(z)$. The triple and higher poles come from contractions between the second term in \G\ and \L. Using the OPE algebra for $N^{mn}$ and multiplying by $\lb_\a(y) \lb_\b r_\g r_\d r_\r(z)$, we obtain
$$
{1\over(y-z)^3} (\lb\g_{mn}\p\th) (r\g^{mpq}r) (\lb\g_{pq}{}^n r) .$$
Factoring out $\lb_\a$ from the first term, $\lb_\b$ from the last term, symmetrizing in $(\a\b)$ and using the identity \Fierz, we obtain that this singularity is proportional to
$$
\lb_\a \p\th^\a (\lb\g_{mnp}r)(r\g^{mnp}r) ,$$
which vanishes because $(\g_{mnp}r)_\a (r\g^{mnp}r)=0$ as consequence of the identity
\eqn\FFi{
\g_m^{\a\d} (\g^m)^{\b\g}=-\ha\g_m^{\a\b}(\g^m)^{\g\d}+{1\over 24}\g_{mnp}^{\a\b}(\g^{mnp})^{\g\d} ,}
which is implied by \Fierz. Consider now $b_1(y)b_2(z)$. The triple and higher poles come from contractions between the second term in \H\ with \K. Using the OPE algebra for $N^{mn}$ and multiplying by $\lb_\a r_\b(y) \lb_\g r_\d r_\r(z)$, we obtain
$$
{1\over(y-z)^3} (\lb\g_m\g_n\p\th) (r\g^{mpq}r) (\lb\g^n{}_{pq} r) .$$
Factoring out $\lb_\a$ from the first term, $\lb_\b$ from the last term, symmetrizing in $(\a\b)$ and using the identity \Fierz, we obtain that this pole vanishes.

At order $r^4$ , we consider the OPE's $b_1(y)b_3(z) - (y \leftrightarrow z)$ and $b_2(y)b_2(z)$. The contractions in $b_1(y)b_3(z)$ do not provide triple or higher order poles. The OPE $b_2(y)b_2(z)$ does contribute. It gives
$$
{1\over(y-z)^3} (r\g_{mnp}r) (r\g^{mnq}r) (\lb \g^p \g_q \g_r \lb) \Pi^r .$$
The term involving two $\lb$ is zero because the pure spinor condition. The product of three gamma matrices can be expressed in terms of $(\lb\g^m\lb)$ plus a term proportional to $(\lb\g^{mnp}\lb)$.

At order $r^5$, the poles com from $b_2(y)b_3(z) - (y \leftrightarrow z)$. The contractions here do not provide triple or higher order poles. It remains the order $r^6$ which come from $b_3(y) b_3(z)$. Using the OPE algebra for $N^{mn}$ we obtain a fourth and a triple poles of the form
$$
{1\over(y-z)^4} [
(r\g_{mnp}r)(\lb\g^{pqr}r)
(r\g^{mns}r)(z)(\lb\g_{spq}r) + (r\g_{mnp}r)(\lb\g^p{}_{qr}r)(r\g^{mp}{}_sr)(\lb\g^{srn}r) ] $$
$$
+{1\over(y-z)^3} [ (r\g_{mnp}r)(\lb\g^{mnq}r)(r\g_{stu}r)(\lb\g^{ps}{}_qr) N^{tu} + \cdots ] .$$
This expression vanishes because $(\lb\g_{mnp}r)(\lb\g^{mnq}r)=0$ and $(\lb\g_{mnp}r)(\lb\g^{mqr}r)=0$ as consequence of \Fierz\ and the pure spinor conditions. Note that the $\cdots$ terms above always involve the combination $(\lb\g_{mnp}r)(\lb\g^{mqr}r)=0$.

Up to now we have proven that the OPE $b(y)b(z)$ has a single pole only in the form \BB\
$$
b(y)b(z)\to{O(z)\over(y-z)} ,$$
where $O$ is annihilated by the BRST operator \Q, has conformal weight $3$ and is primary  according to \totres. It turns out that $O$ can be expanded in powers of $r$ as
\eqn\expansion{O=\O+r_\a\O^\a+r_\a r_\b\O^{\a\b} +\cdots+r_\a r_\b r_\g r_\d r_\r r_\s \O^{\a\b\g\d\r\s} .}
Note that $\O, \dots,  \O^{\a\b\g\d\r\s}$ are fields which depend on $\lb_\a$ and on the minimal variables through the supersymmetric combinations $d_\a, \Pi^m, \p\th^\a$. Note that there is no terms involving derivatives of the pure spinor $\l^\a$ nor the non-minimal pure spinor $r_\a$. In principle it is possible the appearance of such terms in \expansion\ because they could be generated by Taylor expansion of higher poles in the $b(y) b(z)$ OPE. Consider terms with one factor of $r$ in $b(y) b(z)$. As we already noted, the combination $b_0(y) b_1(z) + b_1(y) b_0(z)$ is the relevant one in this case. We see that a triple pole here will be proportional to 
$$
{\lb_\a\over(\l\lb)}(z){\lb_\b r_\g\over(\l\lb)^2}(y) \g_{mnp}^{\b\g} (\g^{mnp} d(y))^\a ,$$
which come from contractions between $\p^2\th^\a$ in $G^\a$ with $(d\g_{mnp}d)$ in $H^{\b\g}$. When we perform a Taylor expansion to get a single pole we note that all the terms involving a factor $\lb_\a\lb_\b$ vanish because the first pure spinor condition in \nonmps\ and the identity $\g_m\g_{npqrs}\g^m=0$. We obtain a single pole contribution of the following type
$$
\lb_\a\p\lb_\b\p r_\g (\g_{mnp})^{\b\g} (\g^{mnp})^{\a\d} .$$
Although this expression does not vanish, we will show that it can be written as an expression without derivatives on $r$. We first note that the $\g$-matrices combination here can be written as \FFi. Then the single-pole contribution becomes proportional to
\eqn\exx{
\lb_\a\p\lb_\b\p r_\g \left( \g_m^{\a\g} (\g^m)^{\b\d} + \ha \g_m^{\b\g} (\g^m)^{\a\d} \right) .}
In the first term we can use $(\lb\g_m\p r)=-(\p\lb\g_m r)$ which is consequence of the second pure spinor condition in \nonmps. For the second term we use that $(\p\lb\g_m\p r)$ is proportional to $(\p^2\lb\g_m r)+(\lb\g_m\p^2 r)$ because of the same pure spinor condition. When we insert this result into \exx, we note that this expression becomes proportional to 
$$
(\p\lb\g_m r)(\g^m \p\lb)^\b + {1\over 4}(\lb\g_m\p^2\lb)(\g^m r)^\b .$$
Similarly all the terms with potential derivatives on $r$ can be shown to be written as expressions without derivatives on $r$ as consequence of the pure spinor conditions \nonmps.

Note also that $N^{mn}$ and $J$ in \bnm\ produce singularities when act on $\l^\a$ of the denominators in \bnm. These contractions are single poles and are all of the form \expansion.

The BRST operator \Q\ can be written as $Q=Q_0+Q_1$, where $Q_0$ is the BRST charge in the minimal pure spinor formalism and $Q_1=\oint \ob^\a r_\a$. The equation $Q O=0$ can also be expanded in powers of $r$. Since $Q_1 r_\a=0$ and $Q_1\lb_\a=r_\a$ we expand as
\eqn\expa{Q O=Q_0\O+(Q_1\O-r_\a Q_0\O^\a)+(-r_\a Q_1\O^\a + r_\a r_\b Q_0 \O^{\a\b})+\cdots+r_\a r_\b r_\g r_\d r_\r r_\s Q_1 \O^{\a\b\g\d\r\s} .}
Here $Q_0\O$ is of order $r^0$, $(Q_1\O-r_\a Q_0\O^\a)$ is of order $r^1$, etc... Now we will argue that $\O$ is zero. The bosonic field $\O$ depends on two factors of $\lb_\a$, and on $\Pi^m, N^{mn}, J, d_\a, \p\th^\a$ and their derivatives such that the combination has conformal dimension three. After discarding the terms which vanish because the pure spinor conditions, a possible $\O$ has the form
\eqn\possomega{\O=A(\lb\g^m\p^2\lb)+B\lb_\a\p\lb_\b \p\th^\a \p\th^\b+C(\lb\g_{mnp}\p\lb)\Pi^m N^{np}}
$$+D(\lb\g_{mnp}\p\lb)(d\g^{mnp}d)
\Pi_m+E\lb_\a\lb_\b\p\th^\a\p^2\th^\b+\cdots ,$$
where $A,B,C,\dots$ are fixed by demanding $Q_0\O=0.$  There are more combinations in $\O$ that we are not writing but, it turns out that BRST invariance does not mix them, then all the possible terms in \possomega\ must be discarded. Therefore, $\O$ is absent in \expa. Now we consider the next term in the $r$ expansion \expa and similarly prove that $Q_0\O^\a=0$. Construct all possible terms which are fermionic and have conformal dimension three. Some possible terms are
\eqn\possomegaa{\O^\a=A(\lb\g_m)^\a (\p^2\lb\g^m d)+B(\lb\g_{mnp}\p\lb)(\g^{mnpqr}d)^\a N_{qr}+\cdots,}
but again all possible combinations has to be discarded because BRST invariance. This line of reasoning continues until we show that $O$ is zero.

\vskip 20pt
{\bf Acknowledgements:} I would like to thank Yuri Aisaka, Nathan Berkovits, William Linch, Brenno Vallilo for useful comments and suggestions. This research was partially financed by FONDECYT project 1061050.

\listrefs

\end